\documentclass[fleqn,usenatbib]{mnras}

\usepackage{newtxtext,newtxmath}
\usepackage[T1]{fontenc}
\usepackage{orcidlink}

\DeclareRobustCommand{\VAN}[3]{#2}
\let\VANthebibliography\thebibliography
\def\thebibliography{\DeclareRobustCommand{\VAN}[3]{##3}\VANthebibliography}

\usepackage{graphicx}	% Including figure files
\usepackage{amsmath}	% Advanced maths commands
\title[FRB selection biases]{The Effects of Selection Biases on the Analysis of Localised Fast Radio Bursts}

\author[Seebeck et al.]{
Jerome~Seebeck,$^1$ 
Vikram~Ravi\orcidlink{0000-0002-7252-5485},$^1$ 
Liam Connor,$^1$
Casey J.~Law\orcidlink{0000-0002-4119-9963},$^1$ 
Dana Simard\orcidlink{0000-0002-8873-8784},$^1$
and Bade D. Uzgil$^1$ \\
$^1$Cahill Center for Astronomy and Astrophysics, MC\,249-17 California Institute of Technology, Pasadena CA 91125, USA. \\
}
\date{Accepted XXX. Received YYY; in original form ZZZ}

\pubyear{2021}

\begin{document}
\label{firstpage}
\pagerange{\pageref{firstpage}--\pageref{lastpage}}
\maketitle

\begin{abstract}

The objects that emit extragalatic fast radio bursts (FRBs) remain unidentified. Studies of the host galaxies and environments of accurately localised ($\lesssim1$\,arcsec) FRBs promise to deliver critical insights into the nature of their progenitors. Here we demonstrate the effects of observational selection biases on analyses of the distributions of FRB host-galaxy properties (including star-formation rate, SFR, and stellar mass, $M_{*}$), and on the distributions of FRB offsets from the centres of their hosts. We consider the effects of ``radio selection'', wherein FRBs with larger dispersion measures and scattering timescales are less likely to be detected, and the effects of ``optical selection'', wherein FRBs with fainter host galaxies are more likely to have unidentified or mis-identified hosts. We develop a plausible, illustrative model for these effects in observations of FRBs and their host galaxies by combining the output catalogues of a semi-analytic galaxy formation model with a recently developed algorithm to associate FRBs with host galaxies (PATH). We find that optical selection biases are most important for the host-galaxy $M_{*}$ and SFR distributions, and that radio selection biases are most important for the distribution of FRB projected physical offsets. For our fiducial simulation of FRBs at $z<0.5$, the selection biases cause the median host-galaxy SFR to be increased by $\sim0.3$\,dex, and the median $M_{*}$ by $\sim0.5$\,dex. The median projected physical offset is increased by $\sim2$\,kpc ($\sim0.25$\,dex). These effects are sufficiently large so as to merit careful consideration in studies of localised FRBs, and our simulations provide a guide towards their mitigation. 

\end{abstract}

\begin{keywords}
galaxies: evolution --- ISM: general --- methods: data analysis --- methods: statistical --- radio continuum: transients --- scattering
\end{keywords}

%%%%%%%%%%%%%%%%%%%%%%%%%%%%%%%%%%%%%%%%%%%%%%%%%%

%%%%%%%%%%%%%%%%% BODY OF PAPER %%%%%%%%%%%%%%%%%%

\section{Introduction}

Substantial insights into novel classes of extragalactic transients are gleaned from the properties of their host galaxies, and their locations within their hosts. For example, a correlation between the locations of long-duration $\gamma$-ray bursts (LGRBs) and the UV light-distributions of their hosts led \citet{bkd02} to associate LGRBs with the deaths of massive stars. A finding of comparable explosion sites between LGRBs and Type Ic core-collapse supernovae (CCSNe) indicated a link between the two classes of explosion \citep{kkp08}, but larger metallicities among the host galaxies of (broad-lined) Ic supernovae and LGRBs suggested that only some of these supernovae produce LGRB emission \citep{jvs+18}.  Like LGRBs, superluminous supernovae (SLSNe) are typically associated with the regions of densest star formation within their host galaxies \citep{lcb+15}, but the typically less massive hosts of SLSNe than LGRBs again hint at distinct progenitor pathways \citep{tp21}. \citet{fb13} showed that short $\gamma$-ray bursts are typically offset from the star formation and stellar mass distributions of their hosts, implicating an old progenitor population that migrates widely from the birth sites. The extreme offsets of Calcium-rich gap transients from the centres of light of their host galaxies present an ongoing puzzle \citep{dkt+20}. 

Studies of the host galaxies and environments of accurately localised fast radio bursts (FRBs) offer perhaps our best opportunity to discern FRB progenitors.\footnote{In this work, we only refer to FRBs with positions determined to better than a few arcseconds as accurately localised.} This is especially the case given the absence of easily observable multiwavelength counterparts to FRBs \citep{crl20} and FRB sources \citep[e.g.,][]{prs}. The first localised repeating FRB source \citep{clw+17} and its recently discovered analogue \citep{nal+21} originate from regions of significant star-formation within dwarf galaxies, and are associated with highly magnetised environments and persistent radio sources. However, the remaining sample of localised repeating FRBs originates in a diverse range of environments, from a globular cluster \citep{kmn+21}, to potentially more prosaic locations and galaxies \citep{mnh+20,hps+20,rll+21,fdl+21,bha+21}. The first localisations of apparently non-repeating FRBs revealed host galaxies that are similar to the Milky Way \citep{bdp+19,rcd+19}, in marked contrast to the host of the first localised repeating FRB \citep{clw+17}. The subsequent sample of localisations of apparently non-repeating FRBs \citep{pmm+19,mpm+20,lbp+20,hps+20,bha+21} also revealed a diverse selection of host galaxies and environments, ranging over a few orders of magnitude in stellar mass ($M_{*}$), star-formation rate (SFR), and projected physical offset from the centres of light \citep{bha+21}. A Hubble Space Telescope study of the environments of a sample of eight FRBs with sub-arcsecond localisations revealed a link with spiral-arm structures, but generally unremarkable underlying stellar-mass and SFR densities \citep{mfs+20}. Comparisons of the host-galaxy properties and projected physical offsets\footnote{We use the term ``projected physical offset'' to refer to the rest-frame distance in the plane of the sky between the transient location and the centre of light of the host galaxy.} of FRBs with populations of other extragalactic transients and compact objects generally indicate similarities with LGRBs and CCSNe, but little else \citep{brd21,bha+21}. The wide diversity in host galaxies and host environments within the FRB population paints an inconsistent picture, possibly hinting at multiple progenitor channels, or possibly implicating selection biases in the observed FRB sample. 

Here we demonstrate the potential effects of observational selection biases on the analysis of FRB host-galaxy properties and the locations of FRBs within their host galaxies. Most transients discovered in optical/IR and high-energy surveys are observable from most locations within their host galaxies \citep[although see, e.g.,][]{jka+19}, and in many cases spectroscopic redshift determination of optical/IR transients/afterglows makes host-galaxy identification unambiguous even at large offsets \citep[e.g.,][]{dkt+20}. FRB observations, on the other hand, are subject to two classes of selection bias:
\begin{enumerate}

\item {\bf Radio selection.} Radio searches for FRBs are often incomplete at large dispersion measures (DMs) and burst temporal widths \citep{kp15,smb+18,2021arXiv210604352T,ckr+21}. For example, the Parkes and ASKAP FRB-search pipelines are insensitive to $>50\%$ of the expected FRB population at ${\rm DM}\gtrsim1000$\,pc\,cm$^{-3}$ \citep{smb+18}, and the CHIME/FRB pipeline is likely insensitive to  $>50\%$ of the expected FRB population for scattering timescales $\tau\gtrsim10$\,ms.\footnote{In this work, all scattering timescales $\tau$ are referred to a frequency of 1\,GHz, assuming a scaling of $\tau\propto\nu^{-4}$.} Regions with higher local SFRs within galaxies are likely viewed through larger columns of interstellar medium (ISM) \citep[e.g.,][]{wfv+11}, which implies larger values of DM and $\tau$ \citep{cws+16}. For example, star formation in the nearby starburst galaxy M82 is mostly contained within a kiloparsec-scale region with a diffuse-ISM electron density of $\sim10-100$\,cm$^{3}$ \citep[e.g.,][]{scb+96}; M82 dominates the SFR within the local few-hundred cubic Mpc \citep{klf+08}. Thus, if FRB progenitors are typically associated with ongoing star formation, most FRBs in the local volume may well have ${\rm DM}\gtrsim10^{4}$\,pc\,cm$^{-3}$ and $\tau\gtrsim10^{4}$\,ms.  

\item {\bf Optical selection.} Localised FRBs must be spatially associated with host galaxies, with the modest assistance of DM information \citep{eb17,path}. In the absence of milliarcsecond-scale localisations, FRB-host associations are often statistical, relying on prior information on the expected hosts. In studies of large FRB samples, incorrectly identified or missing host galaxies may result in biases in the observed distributions of galaxy properties, and FRB locations within their hosts.   

\end{enumerate}

In this manuscript, we develop plausible, illustrative models for the population of FRBs and their host galaxies and environments, and simulate the effects of the radio and optical selections. Given these models, our aims are to (\textit{a}) quantify the  effects of these selection biases on analyses of observed FRB host-galaxy properties and FRB locations within their hosts, and (\textit{b}) identify means by which the selection biases can be more carefully quantified, accounted for and mitigated in future studies of localised FRBs. The main result of this manuscript is in Fig.~\ref{fig:dist}, which shows the simulated effects of the radio and optical selection biases on distributions of FRB host-galaxy properties and projected physical offsets. The manuscript is structured as follows. In \S\ref{sec:pop}, we describe our models for the FRB population, and in \S\ref{sec:radio}--\S\ref{sec:details}, we outline the simulations that we conduct. Our results are presented in \S\ref{sec:results}, and we address our aim (\textit{b}) in \S\ref{sec:discussion}. We conclude in \S\ref{sec:conc}. Throughout we adopt a flat cosmology \citep{pc14} consistent with the galaxy-formation simulations we use, with a Hubble constant of $H_{0}=67.3$\,km\,s$^{-1}$\,Mpc$^{-1}$, $\Omega_{b}=0.0487$, matter-density parameter $\Omega_{M}=0.315$.

\section{Simulating catalogues of localised FRBs}
\label{sec:simulations}

In this section we begin in \S\ref{sec:pop} by describing how we generate lists of FRBs that are then ``observed'' according to the radio and optical selections. We then outline the methods used to simulate the radio and optical selections in \S\ref{sec:radio} and \S\ref{sec:optical} respectively, and finish with a summary of the simulations we conducted in \S\ref{sec:details}. We emphasise that these methods are not intended to ideally match the observed universe, but are instead intended as a plausible demonstration of the effects of selection biases on FRB observations. 

\subsection{Simulating FRBs and host galaxies}
\label{sec:pop}

We simulated the FRB population with the aid of a recent semi-analytic galaxy formation model \citep[L-galaxies;][]{2015MNRAS.451.2663H} applied to dark-matter halo catalogues from the Millennium simulation \citep{2005Natur.435..629S}. Our technique is broadly similar to that used by \citet{sph+20}, who generated their own galaxy catalogues using semi-empirical relations between galaxy and halo properties. Our choice of the 2015 version of the L-galaxies simulation was motivated by the availability of pencil-beam output catalogues of nearly $10^{7}$ galaxies, including simulated spatial information, in cones with $2^{\circ}$ opening angles.\footnote{\url{http://gavo.mpa-garching.mpg.de/MyMillennium/}} We used the simulation corresponding to the \citet{2003MNRAS.344.1000B} stellar populations, and retrieved a catalogue of simulated galaxies, including their right ascensions, declinations, $r$-band magnitudes, redshifts, angular and physical disk sizes, inclinations, SFRs, and values of $M_{*}$. All galaxies with $r$-band magnitudes $m_{r}<40$ were retrieved, totalling 98.7\% of the full catalogue. We then implemented two schemes to choose FRB host galaxies: one where the likelihood of a galaxy hosting an FRB is proportional to SFR, and one where the likelihood is proportional to $M_{*}$. We used a linear probability scale to randomly choose a fixed number of FRB host galaxies (see \S\ref{sec:details} for simulation details) according to each scheme of weighting potential FRB hosts. We also split our simulated FRB host-galaxy catalogues into three redshift bins ($z<0.5$, $0.5\leq z \leq2$, $z>2$), to better understand the effects of cosmic evolution. In this manuscript, we focus on the low-redshift FRBs, which may be more likely to be observed by current surveys \citep{smb+18,rsl+21}, and are not egregiously affected by optical selection effects for reasonable assumptions on optical observations. 

Upon choosing FRB host galaxies, the next critical step in our simulation is to choose locations within the host galaxies. This enables us to simulate the effects of the radio and optical selections. We adopted a simple exponential model for the radial distributions of both stellar-mass and SFR distributions within galaxies \citep[identical to][]{sph+20}, with a fixed ratio of scale radii and heights of 7.3 \citep{2002MNRAS.334..646K}. The distributions are described by the form
\begin{equation}
    P(r,\phi,z) = K\exp(-r/r_{0})\exp(-z/z_{0}),
\end{equation}
where $r$, $\phi$ and $z$ are the radial, azimuthal and vertical coordinates respectively, $r_{0}$ and $z_{0}$ are the scale radii and heights respectively, and $K$ is a normalising constant. The scale radii were derived from the simulation output catalogues. For each simulated FRB, we randomly drew a position within the host galaxy according to the exponential model.

For each FRB, we then estimated the DM (${\rm DM}_{\rm host}$) and scattering timescale $\tau$ contributed by propagation through the host-galaxy ISM. We used the simulated inclinations of each galaxy, together with a randomly chosen azimuthal orientation, to identify the sightline through the galaxy. We modelled the electron-density distribution within each galaxy as identical to the exponential distribution used to model the stellar-mass and SFR density distributions, and integrated the path length through the host galaxy to six scale lengths in both radius and height. We normalised the electron-density distribution through the following calculation. First, we used the \citet{k98} relation between the SFR and H${\rm \alpha}$ luminosity (assuming a dominant ${\rm DM}_{\rm host}$ contribution from $\sim10^{4}$\,K gas and case B recombination):
\begin{equation}
    L_{\rm H\alpha} = 1.26 \times 10^{41} {\rm SFR}\,{\rm erg}\,{\rm s}^{-1},
\end{equation}
where the SFR is specified in solar masses per year. 
We then calculated the ${\rm H\alpha}$ surface brightness of the galaxies, $S_{\rm H\alpha}$, according to their physical scales, which we used to calculate the average emission measure using
\begin{equation}
    {\rm EM}_{\rm host} = 2.75 \left(\frac{S_{\rm H\alpha}}{\rm Rayleigh}\right)\,{\rm pc}\,{\rm cm}^{-6},
\end{equation}
similar to methods in \cite{2017ApJ...834L...7T}. Finally we were able relate ${\rm EM}_{\rm host}$ to the electron-density normalisation by integrating over the exponential model of electron-density distribution. 

\begin{figure*}
    \centering
    \includegraphics[width=1\linewidth]{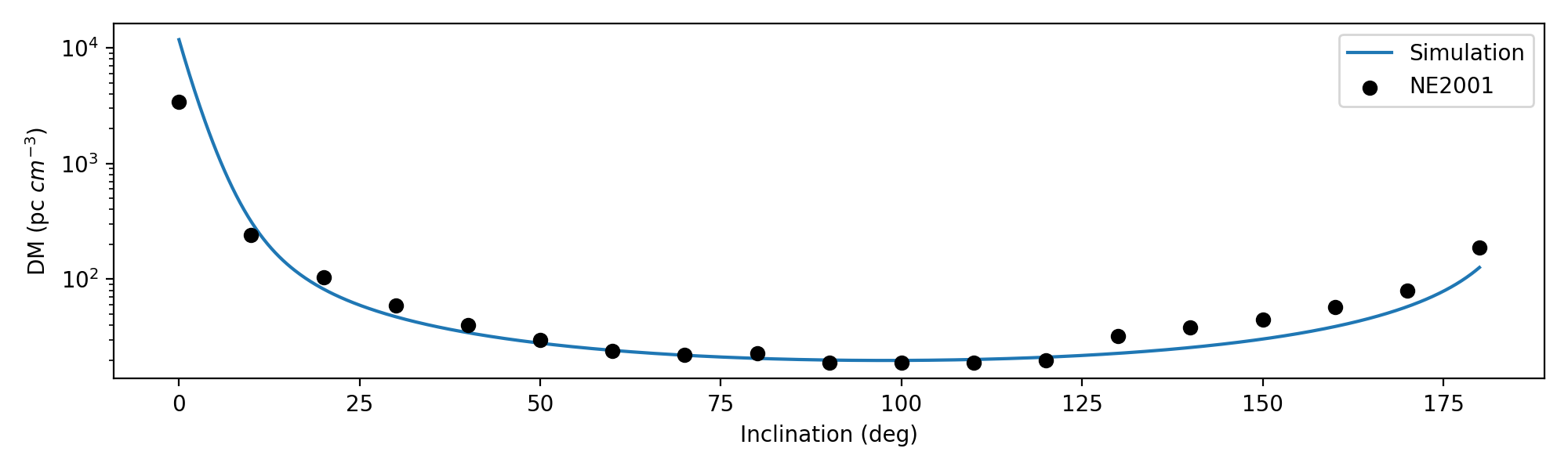}
    \caption{Comparison between our simulated Milky Way DM contribution according to a simple exponential-disk model for the Galactic ISM, and the modelled NE2001 values.}
    \label{fig:fig1}
\end{figure*}

We validated the simulation of ${\rm DM}_{\rm host}$ by modelling the DM contributed by the Milky Way ISM for sightlines to the Earth for different inclinations. We compared our results with the NE2001 Galactic electron-density model \citep{cl02}. We adopted Milky Way parameters summarized in \citet{2015ApJ...806...96L}, including an SFR of 1.65\,$M_{\odot}$\,yr$^{-1}$, a disk scale radius of 2.15\,kpc, and a distance to the Galactic center of 8.33\,kpc. Our results are shown in Fig.~\ref{fig:fig1}; we note that inclinations of $<90^{\circ}$ correspond to Galactic latitudes of $0^{\circ}-90^{\circ}$ for a longitude of $0^{\circ}$, and inclinations of $90^{\circ}-180^{\circ}$ correspond to the same range of Galactic latitudes for a longitude of $180^{\circ}$. Our model approximately reproduces the NE2001 values for most sightlines, although some discrepancies are evident for low and high inclinations, which likely indicate more complex structure close to the Galactic plane.

We simulate the total DM to an FRB as the sum of multiple contributions:
\begin{equation}
    {\rm DM} = {\rm DM}_{\rm MW} + {\rm DM}_{\rm IGM} + \frac{{\rm DM}_{\rm host}}{1+z},
\end{equation}
where ${\rm DM}_{\rm MW}$ is due to the Milky Way ISM and halo, and ${\rm DM}_{\rm IGM}$ is due to the intergalactic medium. For a given redshift, we calculate ${\rm DM}_{\rm IGM}$ using methods found in \cite{2017ApJ...847...22Y}. We add a small amount of variation to ${\rm DM}_{\rm IGM}$ caused by sightline variations, approximately following \citep{dgb+15}. We assign a fixed ${\rm DM}_{\rm MW}=80$\,pc\,cm$^{-3}$, and note that our results are insensitive to this choice. We do not consider any specific contributions from dense ISM immediately surrounding FRB progenitors.  

Finally, we relate the host-galaxy DM to the scattering timescale by using the Milky Way DM-$\tau$ relation from \citet{2021MNRAS.504.1115O}, and assume no contribution from the Milky Way or intervening systems. In addition to the redshift correction \citep[$\tau \propto (1+z)^{-3}$; e.g.,][]{mk13,sr21}, we multiply the host-frame scattering time by a factor of 3 to account for the effects of geometry \citep{cws+16}. For an observer at the edge of the host galaxy, the scattering is distributed between the source and observer, while for an observer in a distant galaxy the scattering all occurs in a thin screen close to the source.

\subsection{Radio selection}
\label{sec:radio}

\begin{figure*}
    \centering
    \includegraphics[width=\textwidth]{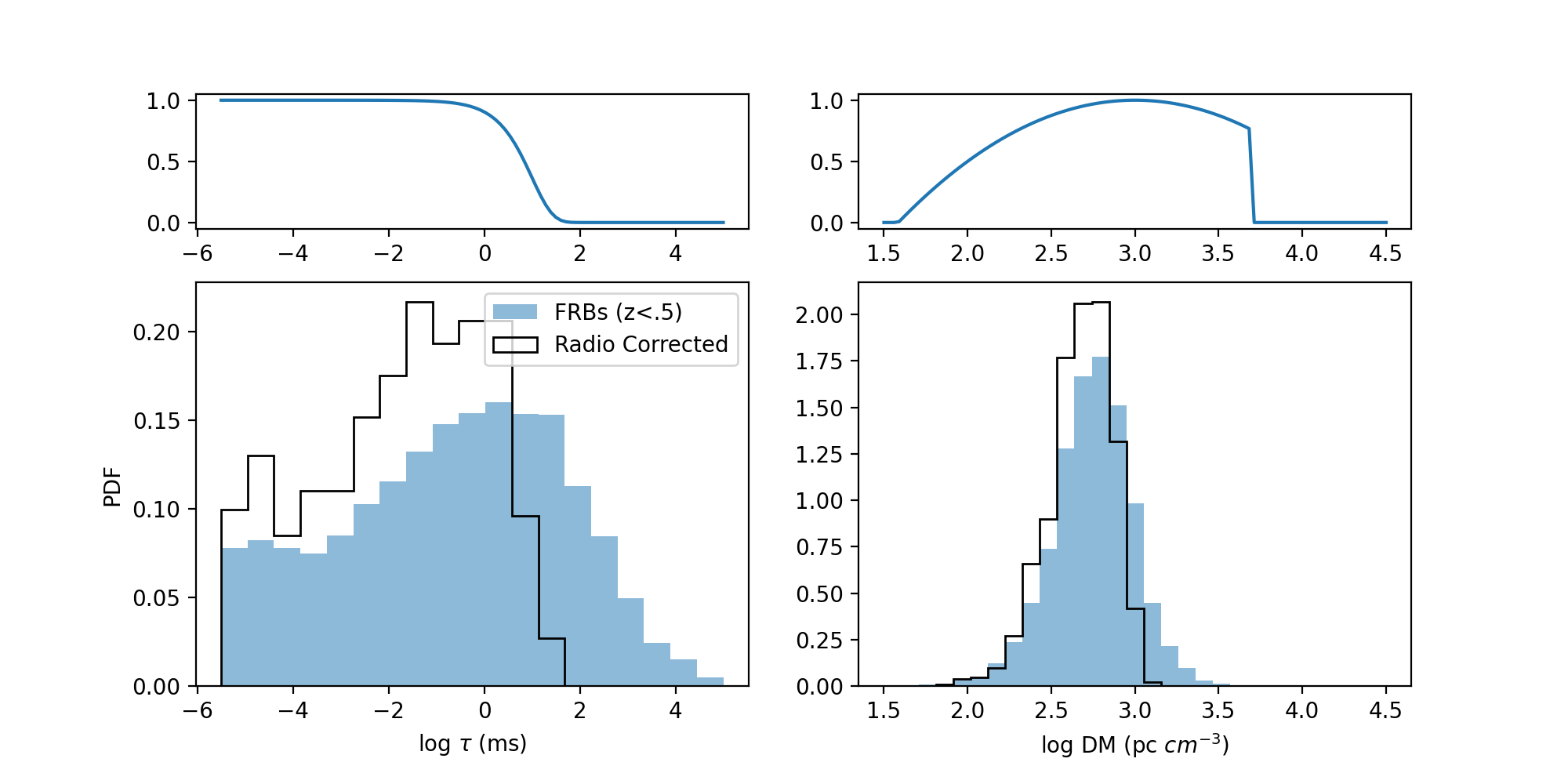}
    \caption{The effects of incompleteness in DM and $\tau$ on the FRB population simulated with the FRB source density proportional to SFR. The top panels show our adopted selection functions, intended to roughly reproduce the selections identified by \citet{2021arXiv210604352T} (their Figures 16 and 17). The bottom panels show normalised histograms of simulated FRBs at $z<0.5$ before (solid blue) and after (black lines) applying the selection functions.}
    \label{fig:fig2}
\end{figure*}

We make use of the careful analysis of survey completeness by \citet{2021arXiv210604352T} to identify representative radio selection functions. We consider incompleteness in DM and $\tau$ using functions that approximately match the CHIME/FRB selections, and include hard cutoffs for ${\rm DM}>5000$\,pc\,cm$^{-3}$ and $\tau>100$\,ms. These functions are shown in the top panels of Fig.~\ref{fig:fig2}. We then use these selection functions to stochastically accept or reject each simulated FRB from the radio-selected samples. 

The bottom panels of Fig.~\ref{fig:fig2} show an example of the radio selection for a simulation of $z<0.5$ FRBs distributed according to SFR. Rough similarity is  evident between our simulations and the CHIME/FRB results \citep[Figures 16 and 17 of][]{2021arXiv210604352T}, both prior and post selection. This is not necessarily expected, given the  $z<0.5$ cutoff of our simulations, although \citet{rsl+21} note that the majority of FRBs observed by CHIME likely originate from these redshifts.

\subsection{Optical selection}
\label{sec:optical}

We now address the observational selection of FRB host galaxies in optical/IR data. In order to simulate host galaxy associations we need a process for using the location of an FRB and properties of nearby galaxies to predict a likely host. Historically chance probability has been used for this exercise \citep{bkd02,eb17}, wherein estimates are made of the false-association likelihoods using simple models for galaxy number counts. For this manuscript we decided to use a new method: Probabilistic Association of Transients to their Hosts \citep[PATH;][]{path}. This is a Bayesian algorithm that calculates the chances of association with specific galaxies, given a set of priors on the nature of FRB progenitors and their locations in galaxies. The priors include information on the nature of the host galaxies, the distributions of projected physical offsets, the probability of the host galaxy being unobserved in the galaxy catalogue, and the FRB localisation ellipse. The ability of PATH to weight probabilities between different galaxies, consider a diverse set of priors, and consider unseen galaxies, makes it a much more robust tool for our work. We used a slightly modified version of PATH which allowed us to consider galaxy priors based on SFR and $M_{*}$; these are optimal for our simulations, because they accurately reflect the means by which we simulate FRB host galaxies. We refer readers to the original PATH paper \citep{path} for more details. 

For each simulated FRB, we used the positions and $r$-band magnitudes of nearby galaxies ($m_{r}$) from the output catalogue of the L-galaxies simulation to attempt to identify the host galaxy using PATH. We used a search radius of 30\,arcsec. We adopted two-dimensional Gaussian FRB-localisation error ellipses with full-width half-maximum (FWHM) of 1\,arcsec; although this is large compared with some existing FRB localisations \citep{hps+20}, it enables us to illustrate the effects of the optical selection. 

\begin{figure*}
    \centering
    \includegraphics[width=1\linewidth,trim={2cm 1.2cm 4.5cm 2cm},clip]{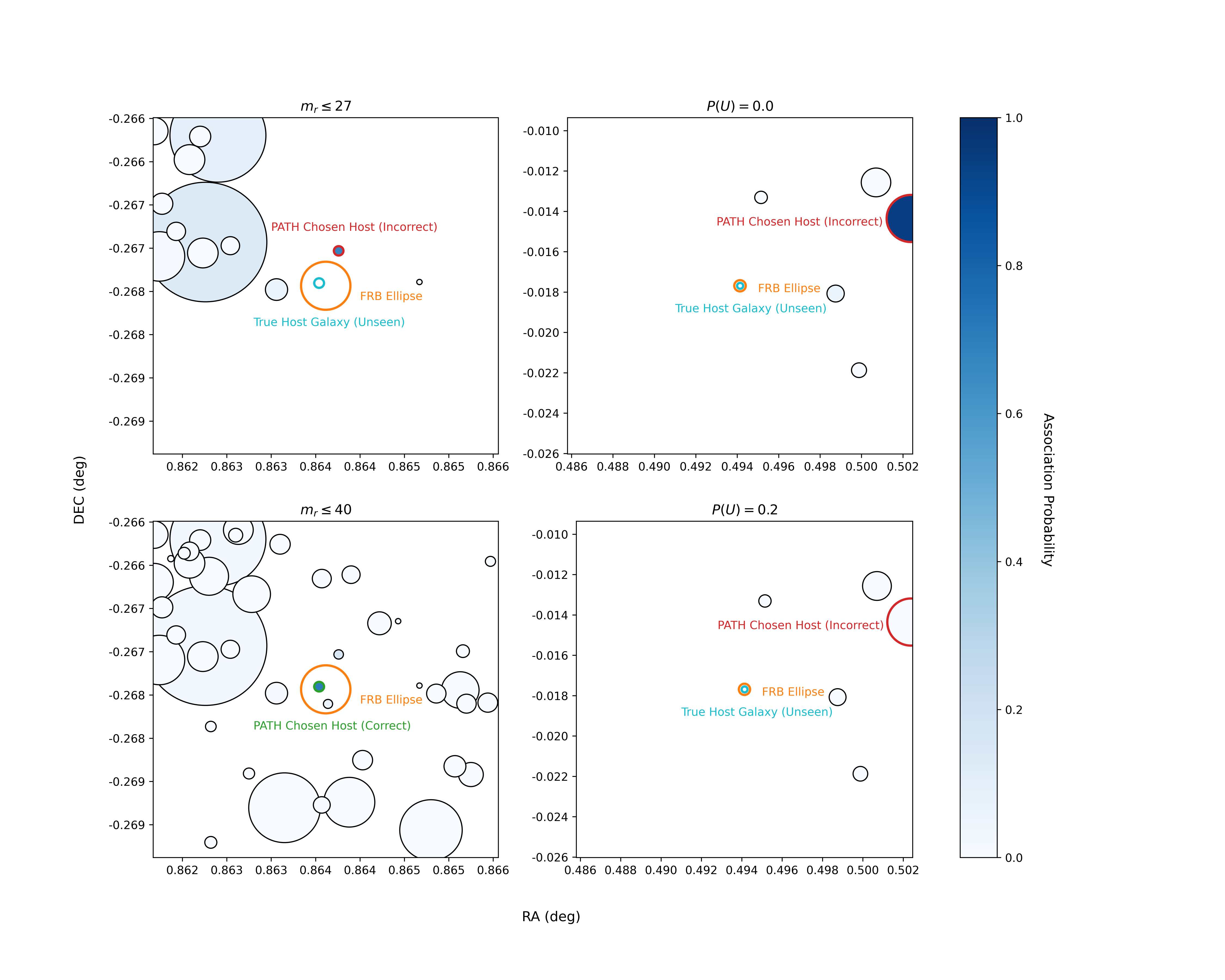}
    \caption{Simulated images of the sky showing possible host galaxies and their association probabilities calculated with PATH, with the FRB localisation ellipses shown in orange. Correctly associated galaxies are outlined in green, incorrectly associated galaxies are outlined in red, and the true hosts are outlined in cyan. \textit{Left column:} misidentification of host due to lack of catalogue depth. In this figure, we consider optical galaxy catalogues with $m_r \leq 27$ and $m_r \leq 40$ (i.e., unlimited). Localisations are shown with the fiducial 1\,arcsec diameter. \textit{Right column:} the top-right panel shows a high-probability association of an incorrect host due to ignoring the possibility that the host galaxy is undetected, with $P(U)=0$. A small increase in the unknown prior to $P(U)=0.2$ results in more accurate association probabilities (bottom right). See text for further detail.}
    \label{fig:fig3}
\end{figure*}

Fig.~\ref{fig:fig3} presents a few PATH runs showing how positive and negative associations arise. The first effect is that with a magnitude limited sample there are occasionally cases in which the true host galaxy is not observed, and thus a negative association is definite. This is shown in the left column of the figure. In the top-left panel, the galaxy catalogue is limited to $m_r<27$, causing the true host galaxy to not be present and thus a nearby galaxy is chosen as the host. In the bottom-left panel, there is no magnitude limit on the sample and thus PATH is able to correctly identify the host with high confidence. With a magnitude limited sample the unknown prior ($P(U)$) becomes necessary for PATH to perform best; this effect is shown in the right column of Fig.~\ref{fig:fig3}, in which a different simulated FRB is used. The top right panel, with $P(U)=0.0$ shows a PATH association in which the true host galaxy is not present and a nearby galaxy is given a very high association probability. The bottom-right panel shows the same run as the top-right panel, but  with $P(U)=0.2$, allowing PATH to lower the association probability for all of the possible host candidates and concluding that the host is most likely unseen. Based on examples like this (see \S\ref{sec:path}), we reject host galaxy associations with probabilities below 0.7, and treat associations with higher probabilities as secure.  This reduces the bias due to low-probability associations when the true host galaxy is undetected in the optical galaxy catalogues. We then have four different types of associations: 
\begin{itemize}
    \item False positives (FP), for which the PATH association was secure but the identified host galaxy was incorrect,
    \item True positives (TP), where we correctly identify the host when accepting the PATH association,
    \item False negatives (FN), where the correct galaxy is identified by PATH with association probability below 0.7, and
    \item True negatives (TN), where an incorrect host galaxy is preferred by PATH with association probability below 0.7.
\end{itemize}

\begin{figure*}
    \centering
    \includegraphics[width=\textwidth]{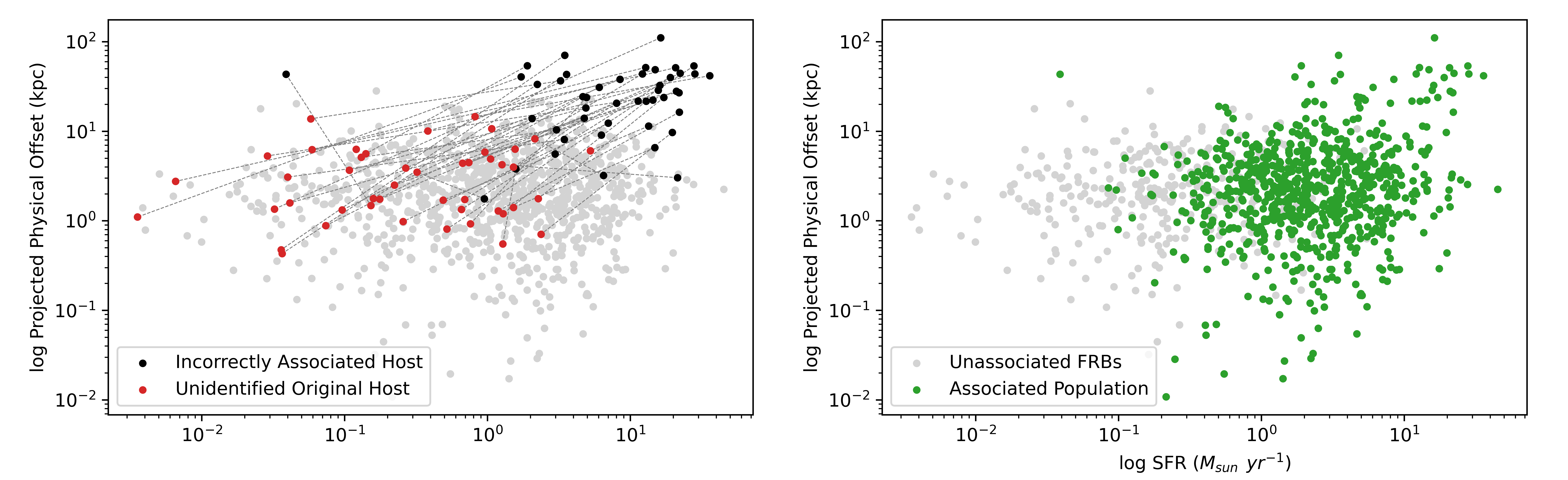}
    \caption{Illustration of the effects of optical selection given a using PATH for a sample of 1,000 radio-selected $z<0.5$ FRBs distributed according to host-galaxy SFR. We assume magnitude-limited optical data with $m_{r}<23.2$. \textit{Left panel:} the effects of falsely associated (FP) hosts are shown. FRBs that have true hosts indicated by red dots are falsely associated with hosts indicated by black dots. Dashed lines join the true and false hosts for each case. The FP hosts are typically more rapidly star forming and located at greater projected physical offsets to the FRBs. Grey dots indicate the remainder of the sample. \textit{Right panel:} the effects of unidentified hosts (both TN and FN) are shown. Green dots represent the host galaxies of FRBs with secure associations in PATH, while grey dots indicate the host galaxies of FRBs for which no secure associations were found. }
    \label{fig:fig4}
\end{figure*}

We considered optical selection biases for magnitude limited samples of possible FRB host galaxies. For the purposes of this paper we used two different magnitude cuts: one at the PanSTARRS $r$-band limit \citep[$m_r=23.2$;][]{ps1}, and the other set at a more optimistic $m_r=25$. This limited the number of galaxies available to PATH, causing potential errors in the FRB host-galaxy identifications. The resulting effects are shown in Fig.~\ref{fig:fig4}. Here we show 1,000 FRBs at $z<0.5$ distributed according to SFR that have already passed the radio selection, but are affected by the optical selection bias for a catalogue limited at $m_{r}=23.2$. In the left panel, we highlight FP host galaxy associations, which typically have higher stellar masses and larger projected physical offsets than the true host galaxies. In the right panel, we compare the properties of the host galaxies identified by PATH (for both TP and FP associations) to the properties of the true host galaxies of FRBs for which PATH finds no secure host galaxy association (TN and FN associations). The associations are biased towards galaxies with higher SFR, but there is negligible bias in the projected physical offsets. The large number of unassociated FRBs is due to our choice of 0.7 for the secure-association cutoff probability, which results in our final catalogue being more accurate but missing some reasonable associations. 

\subsection{Simulation details}
\label{sec:details}

For each redshift range introduced above, we first simulated 10,000 FRBs and host galaxies as described in \S\ref{sec:pop}. We then selected samples of 1,000 FRBs according to the radio selection functions described in \S\ref{sec:radio}. For these samples, we simulated the  procedures used to identify FRB host galaxies as described in \S\ref{sec:optical}. We explored every combination of simulation parameters listed in each column of Table~\ref{tab:1}. The results that we focus on in this manuscript were for both the SFR and $M_{*}$ FRB distributions, the PS1 magnitude cutoff, $z<0.5$, and $P(U)=0.2$. We used an exponential-distribution prior for the offsets from host galaxies in PATH. The choice of the PS1 magnitude cutoff and $z<0.5$ is intended to roughly reproduce results from the ASKAP FRB survey \citep{hps+20} in the absence of deeper optical imaging of each FRB localisation region. 
%The choice of $P(U)=0.2$ is motivated by the considerations in \S\ref{sec:path}. 

\begin{table}
\centering
\caption{Simulation parameters}
\begin{tabular}{ c c c c }
 FRB distribution & $r$-band magnitude cutoff  & Redshift & $P(U)$ \\
 \hline
 $\propto$SFR & 23.2 & $z < 0.5$ & 0\\  
 $\propto M_{*}$ & 25 & $0.5 < z < 2$ & 0.2\\
  & & $z > 2$ & 0.4
\end{tabular}
\label{tab:1}
\end{table}  

\section{Results}
\label{sec:results}

\begin{figure*}
    \centering
    \includegraphics[width=\textwidth]{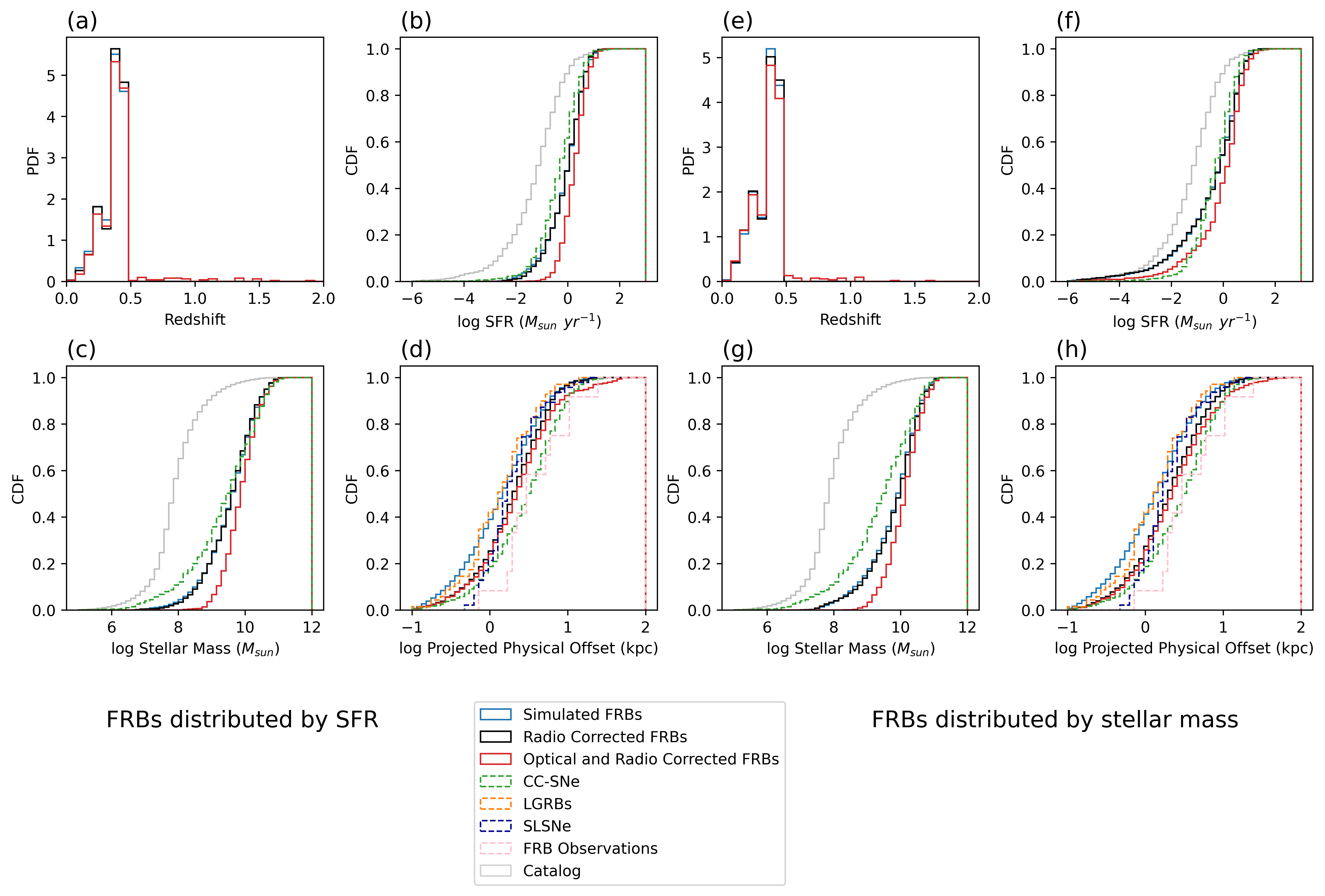}
    \caption{Simulated distributions of FRB host-galaxy parameters and projected physical offsets. We show results for $z<0.5$ FRBs that have been distributed between galaxies based on SFR (panels (a)--(d)) and stellar mass (panels (e)--(h)). We show results prior to any selection bias (light blue solid curves), after simulating radio selection bias (black solid curves), and after simulating both the radio and optical selection bias (red solid curves). The full distribution of the galaxy catalogue (before selecting FRB host galaxies) is shown in grey. We compare our simulation results with distributions of various transients \citep{tp21}, including FRBs \citep{mfs+20}.  Panels (a) and (e) show distributions of FRB redshifts; because we are simulating only $z<0.5$ FRBs, the high-redshift tail in the redshift distribution is due solely to FP associations. Panels (b,f) and (c,g) show cumulative distributions of the host-galaxy SFRs and stellar masses respectively. Panels (d) and (h) show cumulative distributions of FRB projected physical offsets.} 
    \label{fig:dist}
\end{figure*}

Our main results are presented in Fig.~\ref{fig:dist}. The figure demonstrates the effects of the radio and optical selection biases introduced above on simulated FRB populations at $z<0.5$ distributed by both SFR and $M_{*}$. The effects of the selection biases can be summarised as follows:
\begin{enumerate}
    
    \item The radio selection bias negligibly affects the observed redshift distributions (panels (a) and (e)). The optical selection bias introduces a few FP host galaxies at redshifts $z>0.5$. This is caused by the true host galaxies being either un-catalogued or having low association probabilities. 
    \item The radio selection bias negligibly affects the observed distributions of FRB host galaxies in SFR and $M_{*}$ for both means of distributing FRBs among host galaxies (panels (b), (c), (f) and (g)). Only a very small decrease in the number of low-SFR and low-$M_{*}$ hosts is evident due to the radio selection. This is explained by the greater relative probability that in smaller host galaxies the FRBs are found closer to the centers, where the stellar/SFR density is highest, as compared with larger host galaxies. 
    \item The optical selection bias significantly affects the observed distributions of FRB host galaxies in SFR and $M_{*}$. We see increases in the median host SFR of $\sim0.3$\,dex, and increases in the median host $M_{*}$ of $\sim0.5$\,dex. The effect is more pronounced for smaller galaxies. This is explained by the increased likelihood of non-association of smaller host galaxies. 
    \item On the other hand, the distribution of projected physical offsets is most affected by the radio selection (panels (d) and (h)). An increase of $\sim0.2$\,dex in the median projected physical offset is evident. A further increase of $\sim0.05$\,dex is caused by the optical selection. As expected, the selection bias is somewhat more pronounced for smaller offsets. In general, the radio selection functions in DM and $\tau$ cause FRBs located near the centres of their host galaxies to be less likely to be detected, because of the greater path lengths through the host ISM. The effect persists at large offsets because FRBs with larger true offsets are more likely to be in larger host galaxies with larger path lengths through the hosts.  
    \item For large projected physical offsets ($\gtrsim10$\,kpc), the optical selection causes a substantial further shift to larger values in the cumulative distributions. This corresponds to FP associations for FRBs that are further away from their hosts.   
    
\end{enumerate}

In Fig.~\ref{fig:dist}, we also show example distributions, primarily in projected physical offset, of LGRBs, SLSNe, and CCSNe. We also show the empirical distribution of FRB projected physical offsets presented by \citet{mfs+20}. The purpose of this comparison is not to try and reproduce any particular distribution with our simulations; our simulated FRB distributions after accounting for the radio and optical selection biases are not consistent with the empirical distribution. It is however clear that the possible shifts in the distributions caused by the radio and optical selection biases are significant enough that, if unaccounted for, incorrect conclusions may be reached regarding the consistency between the FRB distributions and those of other populations. 

For the higher redshift simulations that we conducted (see Table~\ref{tab:1}), we found that the effects of the selection biases were enhanced. The factors that cause the optical selection biases to be enhanced are clear: in general, fewer galaxies are observed at higher redshifts in a magnitude limited sample, which increases the proportion of FP associations. The radio selection biases are enhanced because the increasing  SFR densities of galaxies, combined with the increased ${\rm DM}_{\rm IGM}$, amplify the effects of the DM and $\tau$ selection functions in our model. 

\section{Discussion}
\label{sec:discussion}

\subsection{Tuning PATH}
\label{sec:path}

The purpose of simulating the optical selection bias was to illustrate the effects of a specific choice of mechanism for associating FRBs to host galaxies. We used PATH in our simulations in order to incorporate the most robust means in the literature to perform this association, and to be able to easily simulate the effects of modifying our assumptions. We simulated  somewhat large FRB localisation regions (FHWM of 1\,arcsec), and shallow optical observations (the PanSTARRS limit of $m_{r}=23.2$) in order to better highlight the effects of the optical selection bias; some FRBs used in current studies \citep[e.g.,][]{bha+21} are better localised, and most have deeper optical observations than are available from PanSTARRS. Current samples are therefore not likely to be subject to optical selection biases as large as those we identify. However, we note that we have used optimal galaxy priors in PATH, and that our models are specific to FRB progenitors distributed among galaxies according to either SFR or $M_{*}$. A substantially different means of distributing even a subset of FRB progenitors (e.g., a preference for old stellar populations, or low metallicities) may result in different optical selection biases. Our results are also important for the case of the larger upcoming FRB-localisation surveys (e.g., DSA-110, CHORD), and for observations with more sensitive FRB-localisation instruments (e.g., MeerKAT, DSA-2000, SKA), where a greater reliance on optical imaging surveys may be necessary, and higher-redshift FRBs may form a larger fraction of the observed population. 

Given specific limitations on available optical data, the optical selection biases are best mitigated with appropriate tuning of the PATH algorithm. In the original PATH paper \citep{path}, the different performances of different choices for the offset and galaxy priors were well examined. However, the effects of varying $P(U)$ were under-explored. The methods presented in this manuscript allow us to explore the effects of varying $P(U)$ within the constraints of our modelling of the FRB population, because we can simulate a diverse population of FRB host galaxies. Here we quantify the performance of PATH in terms of receiver operating characteristic (ROC) curves. 

\begin{figure}
    \centering
    \includegraphics[width=0.5\textwidth,trim={0.5cm 0.6cm 1.9cm 0.8cm},clip]{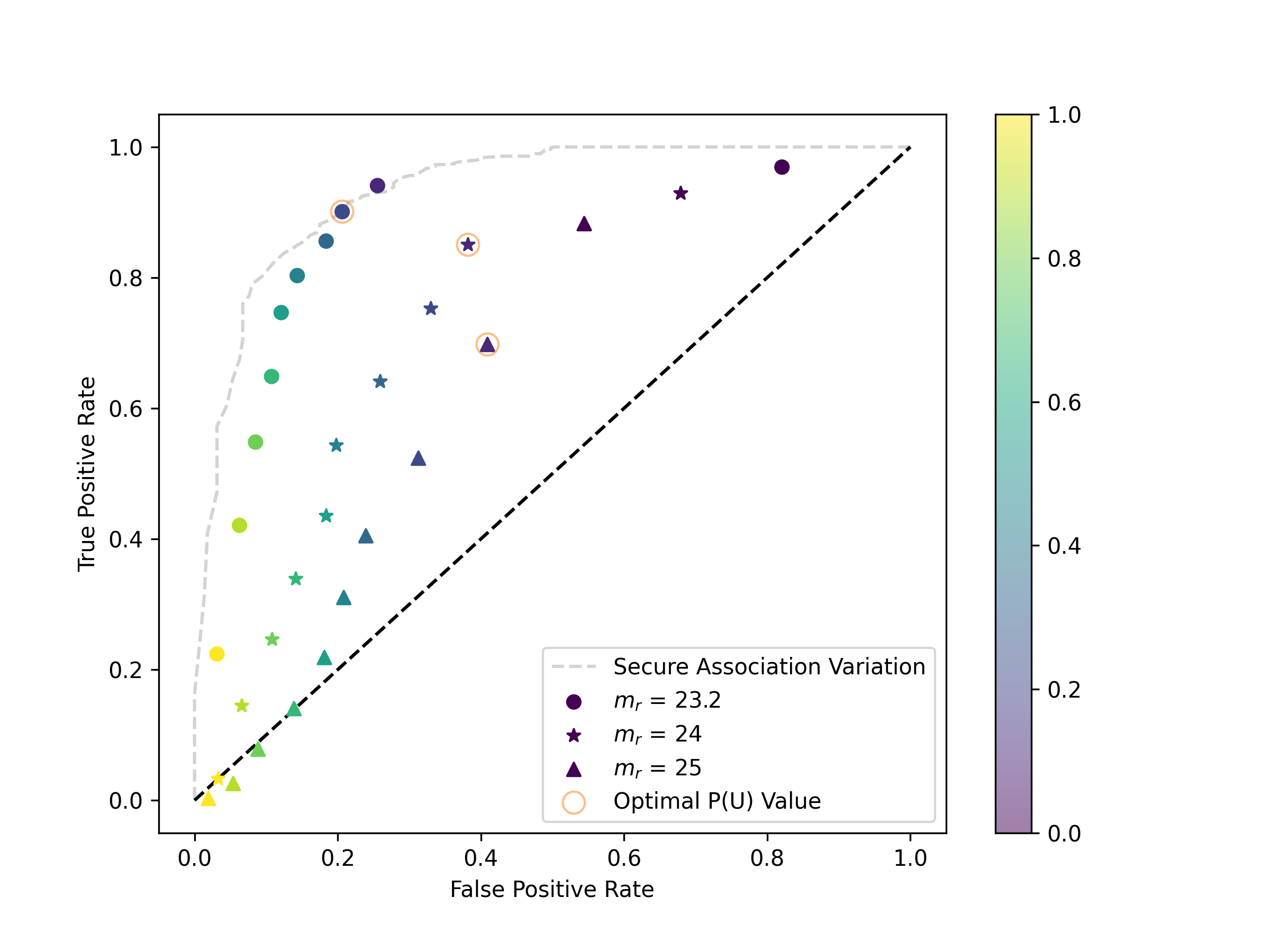}
    \caption{ROC curves indicating the effect of $P(U)$ and the magnitude limit of the optical survey on the TP and FP rates from PATH. All results were derived for 1,000 FRBs distributed according to SFR at $z<0.5$. The symbols show results for different $r$-band magnitude cutoffs, as labelled, where $P(U)$ was varied between 0 and 0.9 according to the colour bar. Optimal values of $P(U)$ chosen according to a minimal distance to the point (0,1) (i.e., ${\rm FP}=0$, ${\rm TP}=1$) are indicated by orange circles. The grey dashed curve shows results for a variation in the secure association threshold between 0 and 1, with $P(U)=0.2$ and a magnitude cutoff of 23.2.}
    \label{fig:fig7}
\end{figure}

In Fig.~\ref{fig:fig7}, we show results for simulated PATH associations of 1,000 $z<0.5$ FRBs distributed among galaxies according to SFR, where we vary $P(U)$ between 0 and 0.9. Results are shown for three different magnitude cutoffs; we fix the secure association cutoff relative probability at 0.7.  We find that the optimal $P(U)$ (i.e. the $P(U)$  for which the TP and FP rates are closest to perfect, with TP=1, FP=0) is characteristically large in our simulations, and as expected appears larger for shallower optical data. In general, we find that the optimal $P(U)$ value is close to the number of true FRB hosts that are below the magnitude cutoff. For example, for $m_{r}=25$, we find that $P(U)=0.7$ is preferred, and 65\% of true host galaxies have $m_{r}>25$. In the figure, we also display an ROC curve for variations in the secure association threshold, where $P(U)$ was fixed at 0.2. The optimal secure association threshold is near 0.7, which motivated this choice elsewhere in the manuscript. 

%We used a $m_r=25$ limited sample of galaxies to associate FRBs and varied P(U) from 0 to 0.9. We can see from the ROC curve that P(U)=0 has a very high TPR and FPR of 0.96 and 0.86 respectively. This is due to the case we saw in Figure~\ref{fig:fig3} in which a P(U) causes PATH to give a high association probability even when the galaxy is likely not the host. In the case  of this example for a a $m_r=25$ it seems as though a P(U) = .7 seems to be the best preforming value. This corresponds closely with the number of FRB true hosts which are not contained in the galaxy catalog which is being searched through 0.65. The plot also contains a ROC curve for a set P(U) value of 0.2 in which we varied the secure association cutoff. The best preforming point is near 0.7 which is what led to the use of that cutoff in other areas of the paper. 

Although the variation of $P(U)$ has no effect on the correctness of PATH, it does change the trustworthiness of the associations. By correctness, we are referrring to the frequency with which the galaxy identified as the most likely host by PATH is the true host galaxy, regardless of the absolute likelihoods. In our brief investigation, deepening the magnitude of the optical galaxy catalogue tends to increase the correctness of the algorithm. However, making the catalogue too deep can cause lower association probabilities because of crowding. This is especially true for FRBs at lower redshifts. For studies of large samples of FRB localisations, we highlight the importance of systematically identifying optimal PATH thresholds and choices of prior, together with simulations like ours of the optical selection biases.  

\subsection{Limitations and extensions of our work}
\label{sec:caveats}

We have presented a specific simulation of the effects of radio and optical selection biases on observations of localised FRBs. We have made specific assumptions regarding the distributions of FRBs in and amongst host galaxies, the selection functions in DM and scattering-timescale $\tau$, the FRB localisation accuracy, and the methods of associating FRBs with host galaxies. Our simulations are also idealised. For example, the methods of simulating values of DM and $\tau$ for different FRBs are based on the assumption of a universal radial profile in stellar mass, SFR, and ionised ISM for all galaxies, and a universal DM-$\tau$ relation based on that of the Milky Way. We also use only a single simulation of the galaxy population. Further, in attempting to demonstrate the radio and optical selection biases for FRB samples that are similar to the currently observed population \citep[e.g.,][]{bha+21}, we simply chose to focus on presenting results for $z<0.5$ FRBs. Variations in our assumptions and modelling procedures will change the results we present (e.g., in Fig.~\ref{fig:dist}) on radio and optical selection biases in the distributions of host-galaxy SFR and $M_{*}$, and the FRB projected physical offsets. 

The optical selection biases can be observationally overcome with FRB localisation accuracies of $\ll1$\,arcsec, coupled with sufficiently deep optical follow-up and optimisation of FRB-galaxy association algorithms like PATH. The latter will be particularly important as FRB surveys reach greater sensitivities. We anticipate that careful modelling of the unseen galaxy population based on established galaxy distributions, together with the consideration of multiple priors on the nature of FRB progenitors, will enable corrections to be made for optical selection bias in distributions of host-galaxy properties. 

On the other hand, the magnitude of radio selection bias on the simulated distribution of FRB projected physical offsets is concerning. This effect is harder to correct for than the optical selection bias. Besides considering different priors on the nature of FRB progenitors, such a correction would require an accurate simulation of the ISM contents in and scattering contributions from FRB host galaxies. A natural extension of our work would be to incorporate cosmological hydrodynamical simulations that accurately resolve the distributions of stars, SFR, and gas within galaxies \citep[e.g.,][]{tng}. An attempt to predict the effects of scattering in host galaxies could be made following the formalism of \citet{sr21}. Such simulations are also likely important to better understand the effects of radio selection biases on other properties of the observed FRB population, such as the relation between DM and $z$ \citep[e.g.,][]{jpm+21}. In our current simulation the effects of the DM and $\tau$ selection functions are comparable; however, observations at substantially higher frequencies than CHIME/FRB may be able to mitigate the effects of scattering. Efforts at measuring survey completeness are also of clear importance \citep[e.g;,][]{2021arXiv210604352T}. We find that the radio selection bias is stronger for higher redshift FRBs, primarily because of the increased DM contributed by the IGM. In general, we caution against the use of distributions of projected physical offsets of FRBs in investigating the nature of their progenitors. Our simulations however suggest that the distributions of FRB host galaxies in SFR and $M_{*}$ are relatively unaffected by the radio selection bias. 

\section{Conclusions}
\label{sec:conc}

We have simulated the effects of selection biases in the distributions of FRB host-galaxy properties, and on the observed positions of FRBs within their hosts. We consider radio selection biases caused by survey incompleteness in DM and scattering timescale $\tau$, and optical selection biases caused by unidentified or misidentified FRB host galaxies (see \S\ref{sec:pop} for details). We considered a specific galaxy formation model \citep[L-galaxies;][]{2015MNRAS.451.2663H} and distributed FRBs in galaxies according to either stellar mass or SFR. We adopted an idealised means of deriving host-galaxy contributions to DM and $\tau$. We simulated the use of a sophisticated algorithm for associating FRBs with host galaxies \citep[PATH;][]{path}. The results of our simulations for the distributions of FRB host galaxies in $M_{*}$ and SFR, and the distributions of projected FRB physical offsets from their host-galaxy centers, are shown in Fig.~\ref{fig:dist}. 

We found that the optical selection biases were most important for the host-galaxy $M_{*}$ and SFR distributions, and that the radio selection biases were most important for the distribution of projected physical offset. For a simulation of $z<0.5$ FRBs, the selection biases cause the median host-galaxy SFR to be increased by $\sim0.3$\,dex, and the median $M_{*}$ by $\sim0.5$\,dex. The median projected physical offset is increased by $\sim2$\,kpc ($\sim0.25$\,dex). The magnitudes of these effects, despite the simplicity of our simulation, motivate their careful consideration in empirical studies of localised FRBs. The optical selection biases are best mitigated with more accurate localisations than we assume (i.e., $<1$\,arcsec), and with deeper optical observations than we assume in our fiducial simulation (i.e., deeper than PanSTARRS). The quality of current observations is such that the optical selection bias is likely minimal in existing analyses \citep[e.g.,][]{bha+21}. We also demonstrate how algorithms such as PATH can be optimised through a survey of their parameter space. However, we caution against interpreting distributions of FRB projected physical offsets without accounting for the radio selection bias. We anticipate that this bias is likely be an important limitation for studies of large samples of FRB host galaxies and environments.

\section*{Acknowledgements}

JS would like to thank the Caltech Summer Undergraduate Research Fellowship program for the opportunity to do this project, and Harold and Mary Zirin and the Zirin family for their generosity in funding the project. This research was supported by the National Science Foundation under grant AST-1836018.



\bibliographystyle{mnras}
\bibliography{frb} 

%%%%%%%%%%%%%%%%%%%%%%%%%%%%%%%%%%%%%%%%%%%%%%%%%%

%%%%%%%%%%%%%%%%% APPENDICES %%%%%%%%%%%%%%%%%%%%%

%\appendix
% FIGURE: Show results while varying along following axes: SFR / stellar mass selection; magnitude cutoff (PS1 \& 25th); redshift bins (z<0.5, 0.5<zM<2, z>2); pU (0, 0.2, 0.4). Panels: Stellar-mass distribution, SFR distribution, offset distribution. 

% Don't change these lines
\bsp	% typesetting comment
\label{lastpage}
\end{document}